\documentclass{llncs}
\usepackage{llncsdoc}

\usepackage[T1]{fontenc}
\usepackage[utf8]{luainputenc}
\usepackage{float}
\usepackage{graphicx}
\usepackage{lmodern}

\makeatletter

\newcommand{\noun}[1]{\textsc{#1}}
\floatstyle{ruled}
\newfloat{algorithm}{tbp}{loa}
\providecommand{\algorithmname}{Algorithm}
\floatname{algorithm}{\protect\algorithmname}

\makeatother

\begin{document}
\title{Exploratory topic modeling\\with distributional semantics}

\author{Samuel R\"onnqvist}

\institute{Turku Centre for Computer Science -- TUCS\\
Department of Information Technologies\\
\AA bo Akademi University, Finland\\
\texttt{sronnqvi@abo.fi}
}

\maketitle


\section*{Abstract}
As we continue to collect and store textual data in a multitude of
domains, we are regularly confronted with material whose largely unknown
thematic structure we want to uncover. With unsupervised, exploratory
analysis, no prior knowledge about the content is required and highly
open-ended tasks can be supported. In the past few years, probabilistic
topic modeling has emerged as a popular approach to this problem.
Nevertheless, the representation of the latent topics as aggregations of semi-coherent terms limits their interpretability and level of detail.

This paper presents an alternative approach to topic modeling
that maps topics as a network for exploration, based on distributional semantics using learned word vectors. From the granular level of terms and their semantic similarity relations global topic structures emerge as clustered regions and gradients of concepts. 
Moreover, the paper discusses the visual interactive representation of the 
topic map\footnote{Topic mapping code and demo available at http://samuel.ronnqvist.fi/topicMap/},
which plays an important role in supporting its exploration.

\begin{keywords}
topic modeling, distributional semantics, visual analytics
\end{keywords}

\setlength{\parskip}{0em}

\section{Introduction}

Following the increase in digitally stored and streamed text, the
interest for computational tools that aid in organizing and understanding
written content at a large scale has soared. Natural language processing
and machine learning techniques demonstrate strength in their feats
of handling the challenging intricacies of human language to extract
information and in their aptitude for scanning big data sets. However,
while we can model what information is likely to be interesting, humans
alone are capable of a deeper understanding that involves evaluating
information against a wide and diverse body of knowledge in nuanced
ways, which motivates a focus on human-computer interaction and visual
analytics in text mining \cite{risch2008text}.

This paper concerns analysis of text by means of \emph{exploratory
topic modeling}, by which I emphasize the exploratory use of models
that convey topic structure. To this end, I put forward a new method
for topic modeling based on distributional semantics using continuous
word vector representations, for the construction of models called \emph{topic maps}. On the one
hand, the focus is set explicitly on unsupervised learning to allow
maximum coverage in terms of domain and language without need for
adaption, while taking advantage of recent advances in word vector
training by neural networks. On the other hand, the role of the human
user is acknowledged as an important part of the analysis process
as the one who understands and explores the modeling results; therefore,
visual interactive presentation is discussed as part of the contribution
alongside map construction and perceived as equally important to exploratory
topic modeling.

Probabilistic topic modeling \cite{blei2012probabilistic} is a family
of machine learning algorithms for uncovering thematic structure in
text documents that are widely used, and applicable both for exploratory
analysis of topics and as a discrete dimensionality reduction method
in support of other learning tasks. Based on co-occurrence of terms
in documents, probabilistic topic modeling extracts a number of latent
topics. In the seminal algorithm, Latent Dirichlet Allocation (LDA),
the number of topics to infer is given as a parameter. Assuming that
each document may discuss a mixture of topics, it attempts to isolate
coherent topics. Each topic is defined as a probability distribution
over terms, where the terms collectively carry the meaning of the
latent topic. While LDA and many of its variations are theoretically
solid and rest on an interpretation-friendly probabilistic basis,
issues of interpretability are nevertheless commonplace and well recognized
\cite{chang2009reading}. First, the unsupervised modeling offers
no guarantees that the topic division is semantically meaningful;
some topics may seem similar and hard to distinguish, whereas others
turn out very specific. These issues may be mitigated by selecting
appropriate parameters, including the number of topics in the case
of LDA. Second, the terms within topics may appear semantically incoherent
and confusing to a human. Various efforts have been made to improve
coherence (e.g., \cite{newman2011improving}\cite{Mimno2011coherence}),
yet for humans to form an understanding of what a topic signifies
based on a set of weighted terms, interpretation inevitably involves
a certain cognitive load, only increased by the iterative task of
contrasting topics against each other to grasp the broader picture. 

Thoughtful visual representation of the topic structure and terms
can ease the task (see, e.g., \cite{ronnqvist2014topic,chuang2012termite}), but I argue
that in many cases it is more meaningful to choose to operate from
the level of individual terms that represent concrete concepts and
their bilateral semantic similarity relations. A discrete division
of topics is practical in many use cases, but is somewhat unnatural
for exploratory purposes, and mere aggregation of terms inevitably
leads toward less interpretable abstractions. Instead, it is more
fitting to allow for a topic structure to emerge as a global property
from the local semantic similarity relations among terms. Such a semantic network
allows the human user to flexibly identify topics as regions through
proper visualization, while the network also supports quantitative
analysis such as community detection \cite{Fortunato201075} (overlapping clustering, which
handles ambiguous terms) to identify discrete topics.

The following section introduces the method for building the semantic
network model, the topic map, whereas Section 3 discusses its visualization,
and Section 4 reports on experiments conducted to demonstrate the
mapping method, followed by some concluding remarks.

\section{Building the topic map}

Distributional semantics models the meanings of words based on their
contexts, namely the surrounding words in a sentence, according to the aphorism
``you shall know a word by the company it keeps'' \cite{firth1968linguistic}.
While modeling has traditionally been based on counting of context words,
recent approaches that work by learning to
predict words instead have been highly successful \cite{baroni2014don}. A popular way of representing the semantics is by vectors, e.g., through projection \cite{Schutze1992dimensions} or later through neural network training \cite{bengio2003nplm}. Lately, Mikolov et al. \cite{mikolov2013efficient} have shown
how neural networks can be efficiently used to train semantic models
based on corpora at the scale of billions of words, in order to achieve
very high semantic accuracy. Their continuous skip-gram model is a
neural network trained to predict context words based on the center
word, using a single hidden layer. Through supervised training, the
network optimizes its hidden layer weights, which results in the learned
array of hidden nodes providing fixed-length vector representations
of word semantics, i.e., word vectors. The word vectors embed words
into a semantic space that supports measuring similarities among words
by their vectors (e.g., by cosine similarity), as well as other vector
arithmetic operations (e.g., addition and subtraction for regularities
prediction).

For the purpose of modeling the general topic composition of corpora,
I use the neural network skip-gram method to model word-level semantic
similarity, and from pairwise relations let the broader topic structure
emerge. Whereas the focus in word vector training generally is to
approximate the semantics of language in general, which can be achieved
by training on large and diverse enough text, the idea is here to
explicitly model the semantics of the language in one's corpus alone.
The model then reflects how words relate in the discourse of the corpus
rather than elsewhere. Thereby, the discrepancies between the word
similarities presented by the model and the observers own, more general
understanding and less data-informed expectation of how the words
relate, constitute telltales of the thematic nature of the underlying
text. (Kulkarni et al. use word vectors accordingly to study linguistic
change in English over time. \cite{kulkarni2014statistically})
For topic modeling to be meaningful, it naturally needs to work for
corpora far smaller than billions of words. As will be demonstrated
in Section 4, skip-gram models can learn usefully accurate word vectors
on much smaller data sets, too.

Apart from semantic similarity, the topic map incorporates term frequencies,
used to represent the prevalence of terms in the corpus, and in combination
with their semantic neighborhood provide a sense of the overall importance
of sections of the map, reflecting the prevalence of specific concepts
or topics. Probabilistic topic modeling similarly uses topic-wise
probability distributions over terms to represent their degree of
importance within the topic.


Using the word vector model and term counts, a semantic network that
constitutes the topic map can be constructed according to Algorithm
1 as described in the following. First, the text of a corpus is processed
and tokenized into meaningful and well normalized terms. Then, the
map is constructed through the following two main steps.


\begin{algorithm}
\noun{\# Word vector training}

model = Word2Vec(\emph{tokens}, vector\_size=\emph{V}, context\_size=\emph{C},
epochs=\emph{E})

\noun{\# Network construction}

\textbf{for} i1 \textbf{in} range(0, \emph{N-1}):

\quad{}\textbf{for} i2 \textbf{in} range(i1+1, \emph{N}):

\quad{}\quad{}t1, t2 = top\_N\_terms{[}i1{]}, top\_N\_terms{[}i2{]}

\quad{}\quad{}net.add\_link(t1, t2, weight=model.similarity(t1,
t2))

\noun{\# Network pruning}

threshold = percentile({[}link.weight \textbf{for} link \textbf{in}
net.links{]}, \emph{P})

\textbf{for} node \textbf{in} net.nodes:

\quad{}cap = sorted(net.links{[}node{]}, key=lambda link: link.weight){[}-1{*}\emph{L}{]}.weight

\quad{}\textbf{for} link \textbf{in} net.links{[}node{]}:

\quad{}\quad{}\textbf{if} link.weight < max(cap, threshold):

\quad{}\quad{}\quad{}net.remove\_link(link)

ws = {[}link.weight \textbf{for} link \textbf{in} net.links{]}

\textbf{for} link \textbf{in} net.links:

\quad{}net.links{[}link{]}.weight = (link.weight-min(ws)) / (max(ws)-min(ws))

\protect\caption{Topic map construction (in: tokens, V, C, E, N, P, L; out: net)}

\end{algorithm}

\textbf{Word vector training.} Given the main parameters, vector size
(\emph{V}) and context size (\emph{C}), word vectors are trained on
term sequences by the method of Mikolov et al. (word2vec). Vector
size determines the dimensionality of the semantic space and is customarily
in the range of 50 to 1000, where higher dimensionality allows for
a finer model given enough data. The size of the word context to consider
is typically about 5-10 words, but for the current task even contexts
up to 25 words have proved satisfactory. Training in multiple epochs
(\emph{E}) (e.g., 3-10) tends to improve the quality of the model
noticeably, especially with little data available.

\textbf{Network construction.} Once the vectors have been
trained, we can use the model to measure similarity of pairs
of terms. The most frequent terms in the corpus are picked for comparison,
preferably excluding stopwords. Typically in the range 100-1000, the
number of unique terms to include (\emph{N}) defines the maximum level
of detail in the topic map and limits the computational complexity
of building it. For each pair, the cosine similarity between their
vectors ($sim(t_{1},t_{2})=\vec{v}(t_{1})\cdot\vec{v}(t_{2})$, with
unit vectors) is computed and stored. 

\textbf{Network pruning.} As only similar terms are meaningful
to relate and as we seek to build a network that is neither too dense
and cluttered nor too sparse and disconnected, the pairs with highest
similarity scores are retained as links between the term nodes. With
varying sizes of the vector and corpora, the similarity scores vary
considerably as well. Thus, filtering of pairs is performed by a threshold
defined as a percentile of all scores stored (\emph{P}), typically
at the 97-99\textsuperscript{th} percentile, which makes the parameter's
effect more stable. Moreover, an upper bound on number of links per
term (\emph{L}) helps reduce cluttering density due to general terms
that may measure as very similar to many terms. Typical cap values
are 8-15 links per term. All links are finally weighted according
to its normalized similarity score, as a standard measure of link strength
($w'\in[0,1]$).

In order to optimize parameter selection the quality of the topic
maps must be evaluated. While the exploratory task ultimately calls
for qualitative evaluation, semantic prediction accuracy will be used
for initial guidance in word vector training, which is the more
computationally demanding step. The evaluation method and data, borrowed
from Mikolov et al., measures syntactic and semantic regularities
such as ``man is to woman as king is to \emph{queen}'', ``Athens
is to Greece as Baghdad is to \emph{Iraq}'' and ``code is to coding
as dance is to \emph{dancing}'', where accuracy in predicting the
last word is evaluated. Measuring how well the model approximates
general English, the relative performance on this task can help to
rule out models that are too simple and produce suboptimal maps because
they lack ability to appropriately model the semantics. The highest
accuracy, however, does not necessarily provide the best topic map,
as its quality relies on a balance between specificity and generality
of its relations. The experiments in Section 4 illustrate this further.

Apart from local link accuracy, the network should ideally show good structure
in terms of how broader
clusters emerge, too. This is highly dependent on both
calibration of the network parameters 
and
how the network
is analyzed. The experiments in this paper focus on visual analysis
based on force-directed layouting, in which case desirable network
structures contain some degree of clustering into coherent and meaningful
regions, without excessive cross-linking between terms in different
clusters to avoid overlaps. The network construction parameters (\emph{P},
\emph{L}, \emph{N}) may be adjusted to optimize the readability of
the map, which in practice can be done instantaneously while visualizing
the network. Hence, optimization of the word vector parameters is
the more cumbersome groundwork that begets good maps, and evaluation
of accuracy helps by reducing the search space.

\setlength{\parskip}{0em}

\section{Visualizing the topic map}

Exploration of complex models such as topic models calls for presentations
that provide as much detail as meaningfully possible. The most information-dense
mode of communication is visualization, whereas interactivity helps
expand the space of information that can be presented intelligibly on a finite
screen. The visual analytics paradigm \cite{keim2008visual} embraces
visual interactive interfaces as they offer a means of communication
that is both rich and reactive, thus, helping users in making sense
of models and data. Visualization of the topic map incorporates Shneiderman's
visual information-seeking mantra, ``overview first, zoom and filter,
then details-on-demand'' \cite{Shneiderman1996}, by providing both
overview of a corpus and a scaffold for exploration of its details.
Visualization of the two main aspects of the map, term frequencies
and word vectors, is discussed in the following, as well as their
combination into a visual topic map.

Among the most popular forms of text visualization are word clouds,
which are simple yet useful. Their main property, representing word
importance by size, is powerful because it utilizes a preattentively
recognized visual variable, i.e., relative word importance is recognized
effortlessly and without requiring focused attention, in parallel
across the field of vision at the early stage of the human visual
system \cite{treisman1986features}. While word clouds have received
some criticism relating to other properties such as the (dis)organization
of words, studies have sought improvement and in terms of readability
evaluated various approaches such as clustered \cite{lohmann2009comparison}
and semantic word clouds \cite{barth2014experimental} that impose
some semantically meaningful organization of words. However, so far
none of the approaches have used distributional semantics, which offers
advantages by being arguably more specific than tried clustering approaches,
and unsupervised as opposed to database approaches.

By contrast, a common approach to visualizing word vectors is to plot
words according to their two-dimensional projections by PCA (or other
multidimensional scaling methods), which achieves a basic form of
semantic organization, albeit easily cluttered at the center. While
word clouds commonly place words as closely as possible, regarding
order or not, projection uses planar distance to communicate the degree
of semantic similarity (a spatial visual metaphor) as well as ordering. Nevertheless, projection
into two dimensions is bound to produce overlap between semantically
unrelated sets of words, which motivates the visualization of semantic
relations by drawn line connections that is more explicit \cite{palmer1994rethinking}.
For visualization of the topic map network, I propose to use a force-directed
layout (a projection method) that optimizes word positions explicitly
based on semantic relations present in the network model, rather than
the whole word vector model. In particular, the D3 force algorithm
\cite{bostock2011d3} is suitable as it can counter overlap of terms
(by node charge) to preserve readability, even when they are densely
connected. It can also run in real time to allow for interactive adjustment
of positions which lets the user explore multiple local optima of
positioning.

The visual topic map lends from word clouds the word sizing relative
to their corpus frequency, and uses force-directed layouting to organize
the map semantically. Drawing the network of words, the strength of
each link is encoded by opacity, which makes more explicit the relative
importance of individual links, and together with the emergent density
of links it provides an aggregate impression of the varying density
of the map.

Interactive exploration of the map is enabled foremost by zoom/pan
capabilities, which in a very direct way allows more terms to be displayed, and highlighting of links of specific terms.
The filtering of terms by frequency can be responsive to the level
of zoom to seamlessly provide more detail on demand. The percentile
filter used to construct the network can be relaxed if the visual
interface can counter the added complexity, and the number
of terms can be increased accordingly. Hence, the scalability of the topic map visualization depends largely on interaction design. The semantic network of frequent
terms also functions as a canvas for other types of information, such as mapping
of local term neighborhoods and relational information, touched upon in Section 5.

\section{Experiments}

The topic map will be demonstrated and tested using two different
corpora. The first corpus is a sample of news articles from Reuters
(U.S. online edition) containing 23k articles and 9.2M words (167k
unique) and the other is a collection of financial patent application
abstracts from the U.S. Patent and Trademark Office comprising 14k
abstracts and 1.7M words (20k unique). In accordance with the exploratory
aim of this article, a topic map is trained and visualized for each
corpus as discussed above, in hope of gaining insight into the thematic
composition of each (see Fig. 1 and 2). The most prevalent terms
representing concrete concepts are displayed and their semantic similarity
relations provide organization that portray topics implicitly as regions
and gradients between them.


\begin{figure}
\begin{centering}
\includegraphics[width=0.98\textwidth]{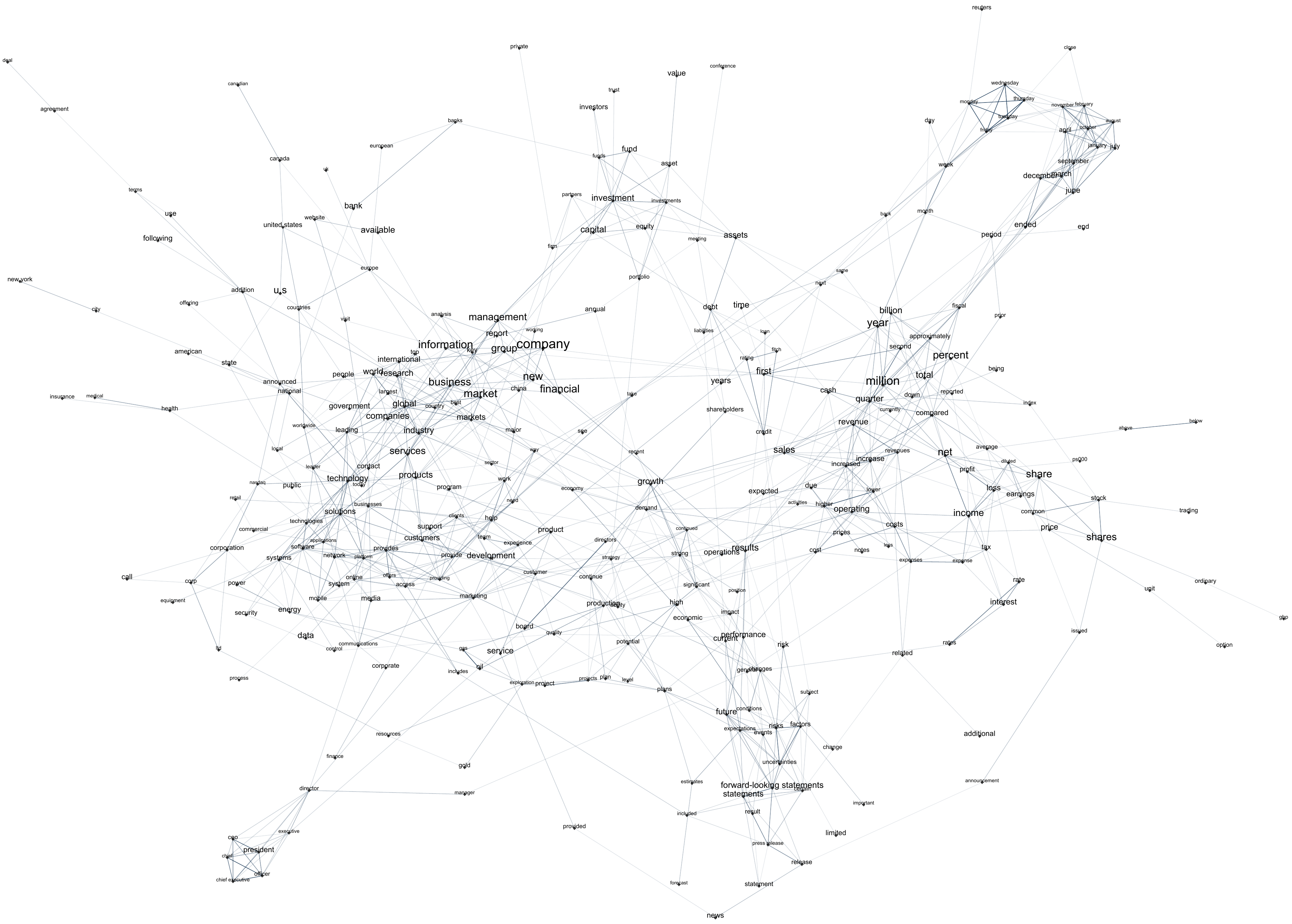}
\par\end{centering}

\protect\caption{Topic map of financial news articles, interactive version available
at: http://samuel.ronnqvist.fi/topicMap/}

\begin{centering}
\bigskip{}

\par\end{centering}

\begin{centering}
\includegraphics[width=0.98\textwidth]{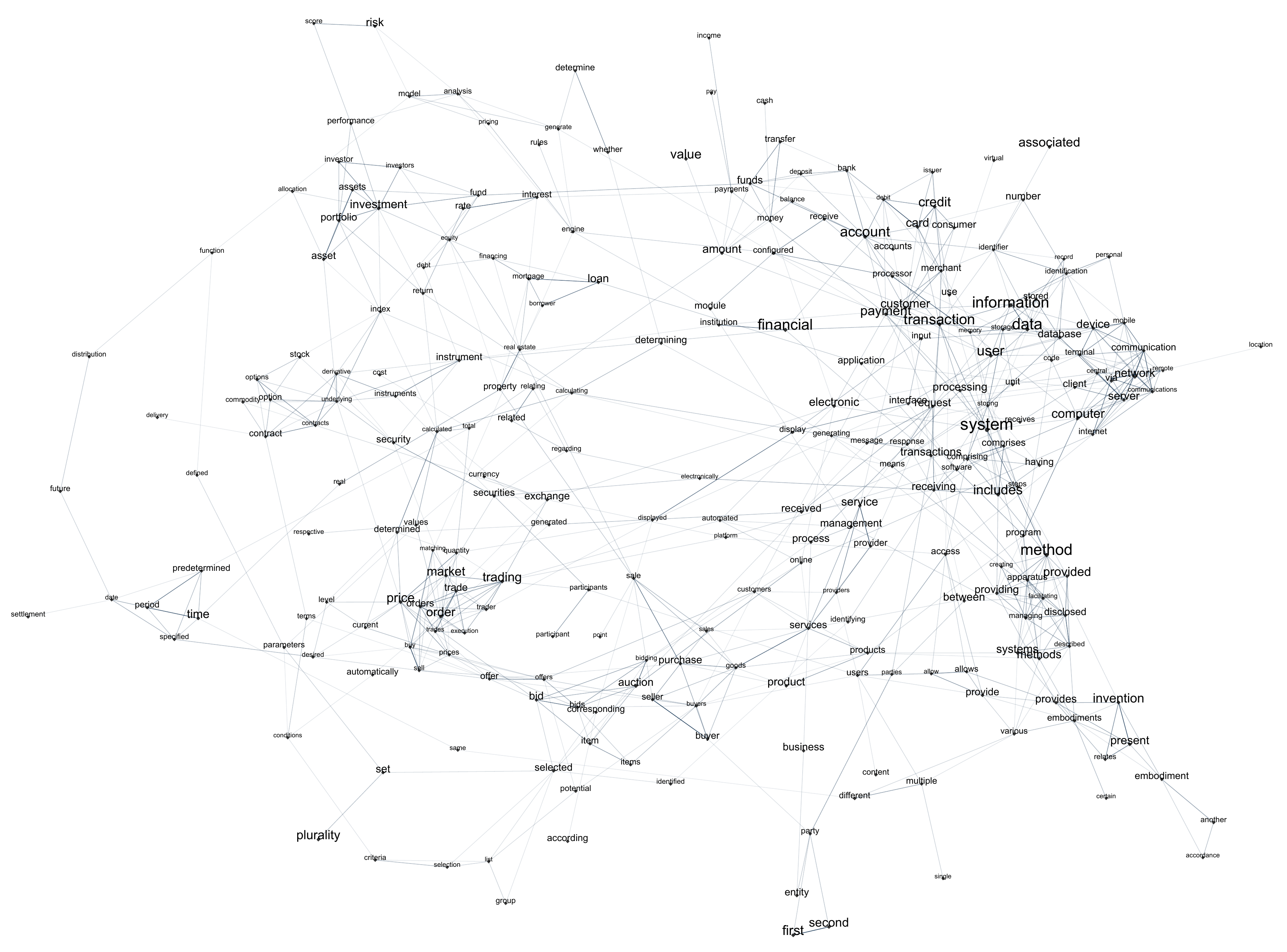}
\par\end{centering}

\protect\caption{Topic map of financial patent abstracts}
\end{figure}

The experiment starts by selecting the parameters for word vector
training, guided by quantitative accuracy and qualitative assessment
of the map (as discussed in Section 2).
Having exhaustively tested various settings, their relationship can be described as follows.
With a fixed context size of 15 for the Reuters corpus,
accuracy reaches a plateau from vector sizes 200 to 500 (on average at 17\%, with \emph{E}=3), decreasing
afterwards. Meanwhile, at a given vector size, accuracy tends to asymptotically
approach a limit with increased context size. Qualitatively, the best
network structure appears to result from settings where accuracy is
close to the limit but context size is kept moderate. 

The experiments
show that the map is surprisingly robust with respect to the training
parameters, producing largely comprehensible results even at vector
sizes of 25 or 600 and context sizes of 5 and 50 respectively. Nevertheless,
the quality of the Reuters map is noticeably best at vector sizes
200-400 and context sizes 10-20, where larger contexts benefit from
larger vectors. Simpler models produce networks with smaller regions
that are tightly clustered, but result in either few or arbitrary
connections between regions, depending on the threshold (\emph{P}).
Networks from complex models have similar problems, although the strong
connections tend to be very specific and semantically accurate, which
explains their good testing performance. 

The qualitatively optimal
models in between strike a balance between, on the one hand, semantic
accuracy that provides a map of meaningful connections and, on the
other hand, generality by connecting parts of the map through more
abstract but still helpfull term relations. Hence, measured accuracy
provides fundamental guidance in learning a model that handles the
language well, but the map then benefits from a slight regularizing
or smoothing effect achieved by using a simpler model than the quantitatively
optimal. While large vectors and contexts combined can achieve maximum
accuracy (about 22\% for the Reuters corpus), it does not seem productive to surpass contexts of about
25 words, and given a limited context size, it is motivated to choose
a vector size towards the beginning of the accuracy plateau. The number
of training epochs has a strong effect on accuracy, e.g., the settings
\emph{V=}400, \emph{C}=15 and \emph{E}=\{1, 3, 5\} give accuracies
7.3, 16.7 and 19.1, but the two latter cases do not show any notable
qualitative difference for the Reuters data.


The topic map in Fig. 1 was produced with the settings \emph{V=}250,
\emph{C}=1\emph{2, E=}5, \emph{N=}500, \emph{P}=.985 and \emph{L=}12 (accuracy 17.6\%, training time 14.5 min on 4 cores).
It depicts the topic landscape of the Reuters financial news corpus
by its most frequent terms excluding stop words (including automatically
detected bi-gram phrases). The similarity threshold set at the 98.5\textsuperscript{th}
percentile provides an appropriate degree of connectivity. The cap
on links per term helps improve readability especially in the dense
region surrounding the terms \emph{business} and \emph{technology}.
The map uncovers an uneven distribution of terms, where smaller concentrations
highlight cliques of terms (e.g., president, ceo, etc. down left)
that represent a rather distinct general concept. Larger concentrated
regions form to highlight a broader topic division of the corpus,
the three main regions broadly reflecting discourse on business-related
activities, realized performance and expected performance.

The map in Fig. 2 similarly illustrates the lay of specific concepts
and more general topics as they occur in the set of patent abstracts.
A few themes can be identified, such as payment systems, telecommunications,
trading, portfolio management and patent-specific language. The map
includes 350 terms and links for the top 2\% most similar pairs. As
the patent corpus is much smaller the vector size was reduced according
to vocabulary size heuristically by $\frac{V_{1}^{2}}{|vocab_{1}|}\approx\frac{V_{2}^{2}}{|vocab_{2}|}$ to \emph{V}=85, context size was kept at 12 not to reduce the already
scarce data and\emph{ }training was run in 10 epochs (training time 3.2 min on 4 cores).


To conclude the evaluation of the generated topic maps, I compare the news corpus against a benchmark obtained by LDA (results for the patent corpus are similar but omitted due to space constraints). The same preprocessing of the text is used as above, and the topics are modeled with standard parameter settings into 8 topics. Each topic is presented by their top-10 terms according to the topic-term probability distributions, as the most direct way of presenting the model. Stop words are excluded to make the results more informative. While several methods have been proposed that rerank terms to better support interpretation of the topics (cf. \cite{wilson2010term,chuang2012termite,ronnqvist2014topic}), no such method seems to have been unanimously or widely adopted. The obtained topics are:

\scalebox{0.75}{
\begin{tabular}{p{1.25cm}p{13cm}}
\tabularnewline
Topic 0: & million,  net,  quarter,  year,  financial,  income,  company,  share,  operating,  total  \tabularnewline

Topic 1: & securities,  class,  relevant,  number,  options,  option,  price,  form,  code,  relevant security \tabularnewline

Topic 2: & company,  shares,  fitch,  fund,  rating,  share,  ratings,  information,  financial,  available  \tabularnewline

Topic 3: & u.s,  bank,  new,  company,  financial,  government,  state,  group,  year,  years  \tabularnewline

Topic 4: & first,  people,  world,  new,  patients,  home,  years,  health,  year,  games  \tabularnewline

Topic 5: & company,  information,  new,  services,  business,  market,  products,  forward-looking statements,  technology,  solutions  \tabularnewline

Topic 6: & q2 2014,  jul amc,  call,  company,  29 jul,  corp,  earnings conf,  jul bmo,  trust,  share \tabularnewline

Topic 7: & percent,  year,  million,  billion,  market,  u.s,  sales,  shares,  growth,  down  \tabularnewline
\tabularnewline
\end{tabular}
}

For some topics it is possible to discern a latent meaning, while others prove hard to interpret. For instance, Topics 0 and 7 appear to relate to realized financial performance, but it is difficult both to form a more detailed explanation of them and to distinguish logically between them. As mentioned in Section 1, recognizing a distinct topic from an aggregate of terms is challenging, as is the task of understanding how multiple topics relate. While the topic map includes many of the same frequent terms, its natural, semantic organization makes it easier to view and grasp the overall topic composition and scope. Local neighborhoods of the map tend to be more coherent than LDA topics, and the relation between different sections of the map is made more explicit. While exploration of LDA topic models can be supported by meaningful presentation (e.g., \cite{chuang2012termite,ronnqvist2014topic}), the topic map's alternative way of approaching topic modeling remains well motivated for exploration.

\section{Discussion}

My aim has been to introduce a new approach of using distributional
semantics, specifically word vectors trained by neural networks, to
explore topics in bodies of text. A problem commonly addressed by
probabilistic topic modeling, this approach sets out to tackle it
with finer granularity, by building a topic map bottom-up from concrete
terms towards general topics, rather than forcing interpretation of
implicit meaning among an explicit, but not necessarily coherent,
set of topic terms. Distributional-semantic modeling provides meaningful
word-to-word similarity relations and organization that is easy to
navigate. In addition, I put forward a visualization design for the map that provides
overview and means for linking to further details, thus supporting
interactive exploration. As a network model, the map also supports
quantitative network analysis, in particular community detection as
a form of second-level clustering to provide explicit topics, which
are useful in some cases. The topic map opens up to a range of possible 
extensions to be explored.

As the map provides a projection of the semantic space of a corpus,
another interesting type of information is the relational, i.e., how
different concepts are referenced together in text. Mapping such relations
onto the topic map may lead to still more informative ways of summarizing
the contents of texts. Document-level co-occurrence of terms used
in probabilistic topic modeling represents a crude way of harnessing
relational information to extract topic information, but it is likely
beneficial to treat distributional word context similarity and word-to-word
co-occurrence as separate aspects that both contribute toward summarizing
the discourse of a corpus. Thus, the approach of constructing a topic
map outlined in this paper should be seen as elementary to future
extensions that among other things include sophisticated analysis
of relations in text and powerful visual interactive interfaces to
make the semantic space and its linked information readily browsable.
The semantic network is the basic data structure, which can be meaningfully
presented in many other ways as well, e.g., using more structured
network layouts or non-graphical representation, possibly emphasizing
search with a completely local focus rather than overview.

Studying immediate neighborhoods of specific terms may in fact be
a desirable mode of exploration, which can be supported in other ways
than described above. Rather than starting from the frequent term
set, terms with the closest vectors can be searched. By recursively
traversing the nearest neighbors of a term, a close-up view of its
semantic context in the corpus is obtainable. 

Vector similarity comparisons can also be performed with compound
vectors that average two or a few word vectors, for instance, as a
way to disambiguate a term (e.g.: \emph{financial} by \emph{financial+group}, \emph{financial+results})
or merge closely related terms (e.g., \emph{customer+customers}). The latter
could be applied to enhance the map by reducing term redundancy and
thereby visual clutter, while joining their term counts. Another way
to generalize across terms would be to smooth term counts to some
extent among direct neighbors, in order to make the representation of 
prevalence of regions more congruent.

In this paper, word vectors and term frequencies were obtained from
the same set of text, which may lead to problems of accuracy for the
study of smaller sets of text (e.g., in the order of 10-100k rather
than 1M words). It is possible to separate these, letting the word
vectors be trained on a larger background corpus while counting terms
on a smaller foreground set, as long as they are related in nature.
For instance, the background corpus may consist of text from a single
source over a certain period of time, while texts from smaller intervals
during that period would be used as foreground corpora to allow for
more specific study of varying term prevalence over time, still benefiting
from a more robust semantic model.

As efficient word vector training with neural networks has opened
up many new possibilities in natural language processing, I hope to
introduce it for the purpose of exploring topics in masses of text
by proposing a methodology for building and visualizing topic maps.
Unsupervised word-level modeling of semantics offers very flexible 
and detailed means for analysis that deserve further study.
The concluding discussion has outlined a few interesting future directions, and ultimately the utility of topic maps and their visual
representations should be tested by how they support users' understanding
in a variety of real-world settings.


\bibliographystyle{plain}
\bibliography{references}

\begin{thebibliography}{10}

\bibitem{baroni2014don}
Marco Baroni, Georgiana Dinu, and Germ{\'a}n Kruszewski.
\newblock Don't count, predict! a systematic comparison of context-counting vs.
  context-predicting semantic vectors.
\newblock In {\em Proceedings of the 52nd Annual Meeting of the Association for
  Computational Linguistics}, volume~1, pages 238--247, 2014.

\bibitem{barth2014experimental}
Lukas Barth, Stephen~G Kobourov, and Sergey Pupyrev.
\newblock Experimental comparison of semantic word clouds.
\newblock In {\em Experimental Algorithms}, pages 247--258. Springer, 2014.

\bibitem{bengio2003nplm}
Yoshua Bengio, R{\'e}jean Ducharme, Pascal Vincent, and Christian Janvin.
\newblock A neural probabilistic language model.
\newblock {\em J. Mach. Learn. Res.}, 3:1137--1155, March 2003.

\bibitem{blei2012probabilistic}
David~M Blei.
\newblock Probabilistic topic models.
\newblock {\em Communications of the ACM}, 55(4):77--84, 2012.

\bibitem{bostock2011d3}
Michael Bostock, Vadim Ogievetsky, and Jeffrey Heer.
\newblock D3: Data-driven documents.
\newblock {\em IEEE Trans. Visualization \& Comp. Graphics (Proc. InfoVis)},
  2011.

\bibitem{chang2009reading}
Jonathan Chang, Sean Gerrish, Chong Wang, Jordan~L Boyd-graber, and David~M
  Blei.
\newblock Reading tea leaves: How humans interpret topic models.
\newblock In {\em Advances in neural information processing systems}, pages
  288--296, 2009.

\bibitem{chuang2012termite}
Jason Chuang, Christopher~D. Manning, and Jeffrey Heer.
\newblock Termite: Visualization techniques for assessing textual topic models.
\newblock In {\em Advanced Visual Interfaces}, 2012.

\bibitem{firth1968linguistic}
John Firth.
\newblock {\em Studies in Linguistic Analysis}, chapter A Synopsis of
  Linguistic Theory 1930-1955, pages 1--32.
\newblock Philological Society, Oxford, 1968.

\bibitem{Fortunato201075}
Santo Fortunato.
\newblock Community detection in graphs.
\newblock {\em Physics Reports}, 486(3–5):75 -- 174, 2010.

\bibitem{keim2008visual}
Daniel~A Keim, Florian Mansmann, J{\"o}rn Schneidewind, Jim Thomas, and Hartmut
  Ziegler.
\newblock {\em Visual analytics: Scope and challenges}.
\newblock Springer, 2008.

\bibitem{kulkarni2014statistically}
Vivek Kulkarni, Rami Al-Rfou, Bryan Perozzi, and Steven Skiena.
\newblock Statistically significant detection of linguistic change.
\newblock {\em arXiv preprint arXiv:1411.3315}, 2014.

\bibitem{lohmann2009comparison}
Steffen Lohmann, J{\"u}rgen Ziegler, and Lena Tetzlaff.
\newblock Comparison of tag cloud layouts: Task-related performance and visual
  exploration.
\newblock In {\em Human-Computer Interaction--INTERACT 2009}, pages 392--404,
  2009.

\bibitem{mikolov2013efficient}
Tomas Mikolov, Kai Chen, Greg Corrado, and Jeffrey Dean.
\newblock Efficient estimation of word representations in vector space.
\newblock In {\em Proceedings of Workshop at International Conference on
  Learning Representations}, 2013.

\bibitem{Mimno2011coherence}
David Mimno, Hanna~M. Wallach, Edmund Talley, Miriam Leenders, and Andrew
  McCallum.
\newblock Optimizing semantic coherence in topic models.
\newblock In {\em Proceedings of the Conference on Empirical Methods in Natural
  Language Processing}, EMNLP '11, pages 262--272, 2011.

\bibitem{newman2011improving}
David Newman, Edwin~V Bonilla, and Wray Buntine.
\newblock Improving topic coherence with regularized topic models.
\newblock In {\em Advances in Neural Information Processing Systems 24}, pages
  496--504, 2011.

\bibitem{palmer1994rethinking}
Stephen Palmer and Irvin Rock.
\newblock Rethinking perceptual organization: The role of uniform
  connectedness.
\newblock {\em Psychonomic Bulletin \& Review}, 1(1):29--55, 1994.

\bibitem{risch2008text}
John Risch, Anne Kao, Stephen~R Poteet, and Y-J~Jason Wu.
\newblock Text visualization for visual text analytics.
\newblock In {\em Visual data mining}, pages 154--171. Springer, 2008.

\bibitem{ronnqvist2014topic}
Samuel R\"onnqvist, Xiaolu Wang, and Peter Sarlin.
\newblock Interactive visual exploration of topic models using graphs.
\newblock In {\em Eurographics Conference on Visualization (EuroVis)}, 2014.

\bibitem{Schutze1992dimensions}
H.~Sch\"{u}tze.
\newblock Dimensions of meaning.
\newblock In {\em Proceedings of the 1992 ACM/IEEE Conference on
  Supercomputing}, Supercomputing '92, pages 787--796, 1992.

\bibitem{Shneiderman1996}
Ben Shneiderman.
\newblock The eyes have it: A task by data type taxonomy for information
  visualizations.
\newblock In {\em Proceedings of the IEEE Symposium on Visual Languages}, pages
  336--343, 1996.

\bibitem{treisman1986features}
Anne Treisman.
\newblock Features and objects in visual processing.
\newblock {\em Scientific American}, 255(5):114--125, 1986.

\bibitem{wilson2010term}
Andrew~T Wilson and Peter~A Chew.
\newblock Term weighting schemes for latent dirichlet allocation.
\newblock In {\em Human Language Technologies: The 11th Annual Conference of
  the North American Chapter of the Association for Computational Linguistics},
  pages 465--473, 2010.

\end{thebibliography}

\end{document}